\documentclass{aa520}
\usepackage{graphicx}
\usepackage{txfonts}
\newcounter{Rco}
\newcommand{\Ionst}[1]{\setcounter{Rco}{#1}\Roman{Rco}}
\newcommand{\Ion}[2]{\mbox{#1\ {\scriptsize\Ionst{#2}}}}
\newcommand{\Ionw}[3]{\mbox{#1\ {\scriptsize\Ionst{#2}}~$\lambda\,#3$\AA}}
\newcommand{\Ionww}[3]{\mbox{#1\ {\scriptsize\Ionst{#2}}~$\lambda\lambda\,#3$\AA}}
\newcommand{\ea}{et\,al\@.}
\newcommand{\logg}{\mbox{$\log g$}}
\newcommand{\loggw}[1]{\mbox{$\log g\hspace{-0.5mm} =\hspace{-0.5mm}  #1$}}

\newcommand{\ab}[1]{\mbox{Fig.\,\ref{#1}}}
\newcommand{\sA}[1]{\mbox{(Fig.\,\ref{#1})}} 
\newcommand{\ratio}[2]{\mbox{$n_{\rm #1}/n_{\rm #2}$}}
\newcommand{\ratiow}[3]{\mbox{$n_{\rm #1}/n_{\rm #2}\hspace{-0.5mm} = \hspace{-0.5mm} #3$}}
\newcommand{\se}[1]{\mbox{Sect.\,\ref{#1}}}  
\newcommand{\sK}[1]{\mbox{(Sect.\,\ref{#1})}}
\newcommand{\spm}{\mbox{\raisebox{0.20em}{{\tiny \hspace{0.2mm}\mbox{$\pm$}\hspace{0.2mm}}}}}
\newcommand{\ta}[1]{\mbox{Table\,\ref{#1}}}
\newcommand{\sT}[1]{\mbox{(Table\,\ref{#1})}}
\newcommand{\Teff}{\mbox{$T_\mathrm{eff}$}}
\newcommand{\Teffw}[1]{\mbox{$\Teff\hspace{-0.5mm} =\hspace{-0.5mm} #1 \mathrm{kK}$}}
\begin{document}
   \title
   {The rotational velocity of the sdOB primary of the eclipsing binary system
    \object{LB\,3459} (\object{AA\,Dor})\thanks
    {Based on observations made with ESO Telescopes at the Paranal Observatories under
    programme ID 66.D-1800.
    }
   }

   \author{T\@. Rauch$^{1, 2}$ \and K\@. Werner$^2$}
   \offprints{T\@. Rauch}
   \mail{Thomas.Rauch@sternwarte.uni-erlangen.de}
 
   \institute
    {Dr.-Remeis-Sternwarte, Sternwartstra\ss e 7, D-96049 Bamberg, Germany
    \and
     Institut f\"ur Astronomie und Astrophysik, Abteilung Astronomie, Sand 1, D-72076 T\"ubingen, Germany}
 
    \date{Received 8 November 2002 / Accepted 17 December 2002}

   \authorrunning{T\@. Rauch and K\@. Werner}
%
   \abstract{We present an analysis of the rotational velocity of the primary of \object{LB\,3459} based on 107 new
             high-resolution and high-S/N ESO VLT UVES spectra. 105 of them cover a complete orbital period 
             (0.26\,d) of this binary system.
             We have determined 
             an orbital period of $P=22\,600.702\spm 0.005\,\mathrm{sec}$, 
             a radial velocity amplitude of $A_1=39.19\spm 0.05\,\mathrm{km/sec}$, and
             $T_0 = 2451917.152690\spm 0.000005$. 
             From simulations of the \Ionw{He}{2}{4686} line profile (based on NLTE model atmosphere
             calculations), we derive $v_\mathrm{rot} = 47\spm 5\,\mathrm{km/sec}$. 

             We present an animation which shows the orbital movement of the binary system,
             its synthetic lightcurve,  
             and compares the phase-dependent variation of the predicted with the 
             observed \Ionw{He}{2}{4686} line profile.

             The radius of the cool component is almost the same size like Jupiter but its mass is about 70 times higher
             than Jupiter's mass. Thus, from its present mass ($M_2 = 0.066\,\mathrm{M_\odot}$), 
             the secondary of \object{LB\,3459}
             lies formally within the brown-dwarf mass range ($0.013 - 0.08\,\mathrm{M_\odot}$).
             It might be a former planet which has survived the previous common-envelope
             phase and even has gained mass.
        }
   \maketitle

\section{Introduction} 
\label{int} 

\object{LB\,3459} is a close binary ($P = 0.26\,\mathrm{d}$) consisting of a sdOB primary star and
an unseen secondary with an extraordinary small mass. Based on the
assumption that the primary component has a mass of $M_1 =
0.5\,\mathrm{M_\odot}$, Hilditch et\,al\@. (1996, hereafter {\sc HHH}) 
found that the cool secondary with $M_2 = 0.086\,\mathrm{M_\odot}$ is in
excellent agreement with lowest mass ZAMS models of Dorman et\,al\@. (1989).

\begin{figure}[ht]
  \resizebox{\hsize}{!}{\includegraphics{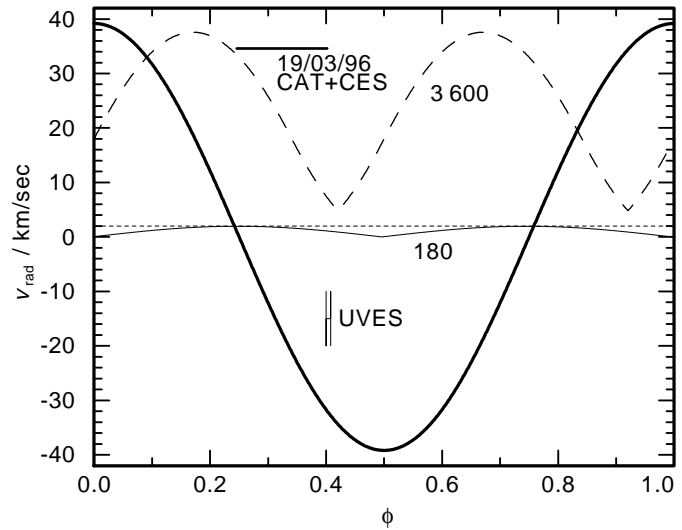}}
  \caption[]{The radial-velocity curve of \object{LB\,3459} 
             (thick line). 
             The phase is given with respect to
             the maximum positive velocity. The 
             {thin and dashed}
             curves indicate the maximum
             velocity coverage during exposures of 180 and 3\,600\,sec, respectively, starting
             at a given phase. 
             The phase coverage of one
             of our UVES spectra and 
             of one CES 
             spectrum used by Rauch (2000) are indicated.
            }
  \label{dvrad}
\end{figure}

\begin{figure*}[ht]
  \resizebox{\hsize}{!}{\includegraphics{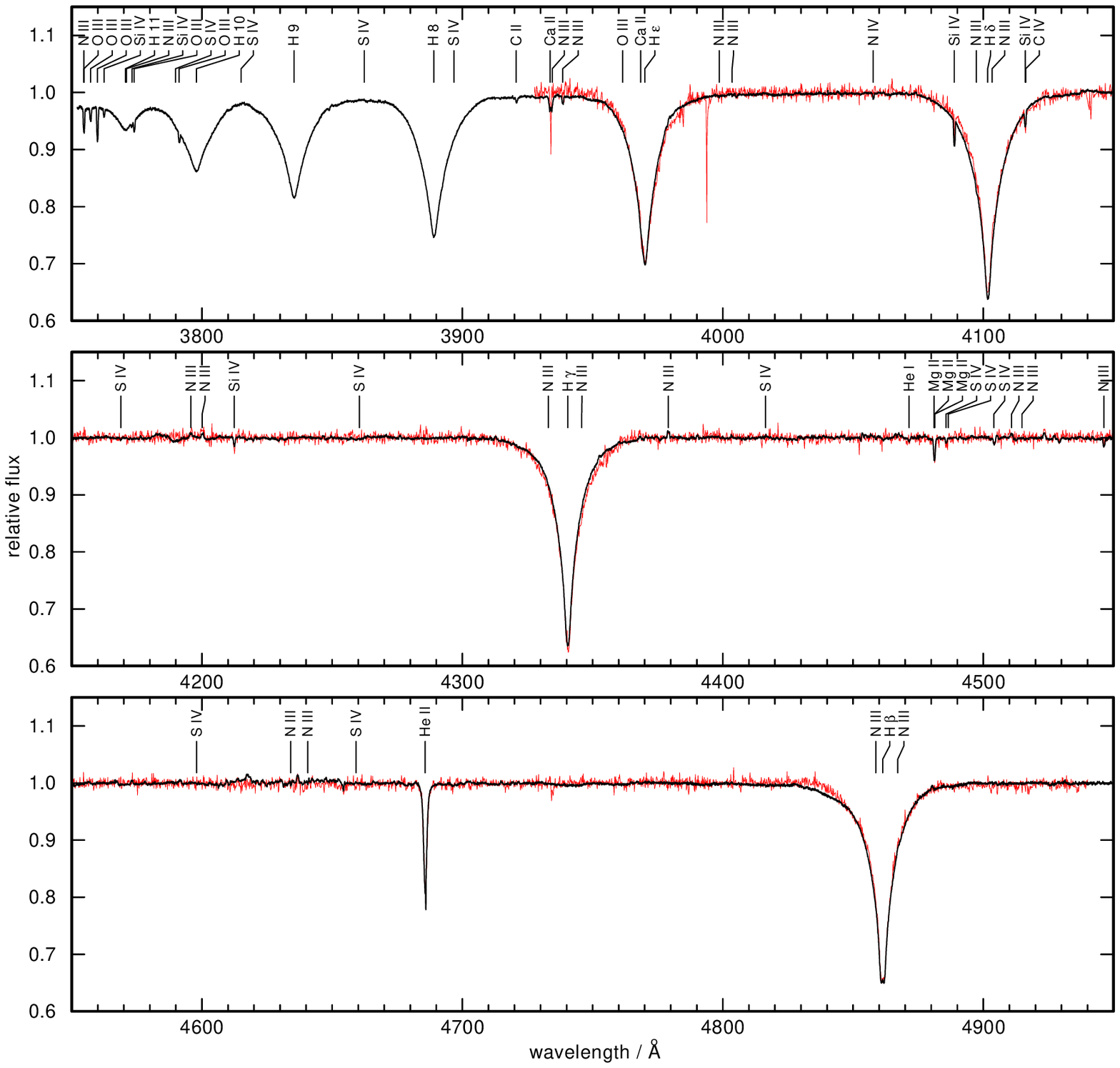}}
  \caption[]{Comparison of the co-added UVES spectrum to a CASPEC spectrum (thin line, 
             observed by U.\,Heber 1985). Identified lines are marked.
            }
  \label{coadd}
\end{figure*}

\begin{figure*}[ht]
  \resizebox{\hsize}{!}{\includegraphics{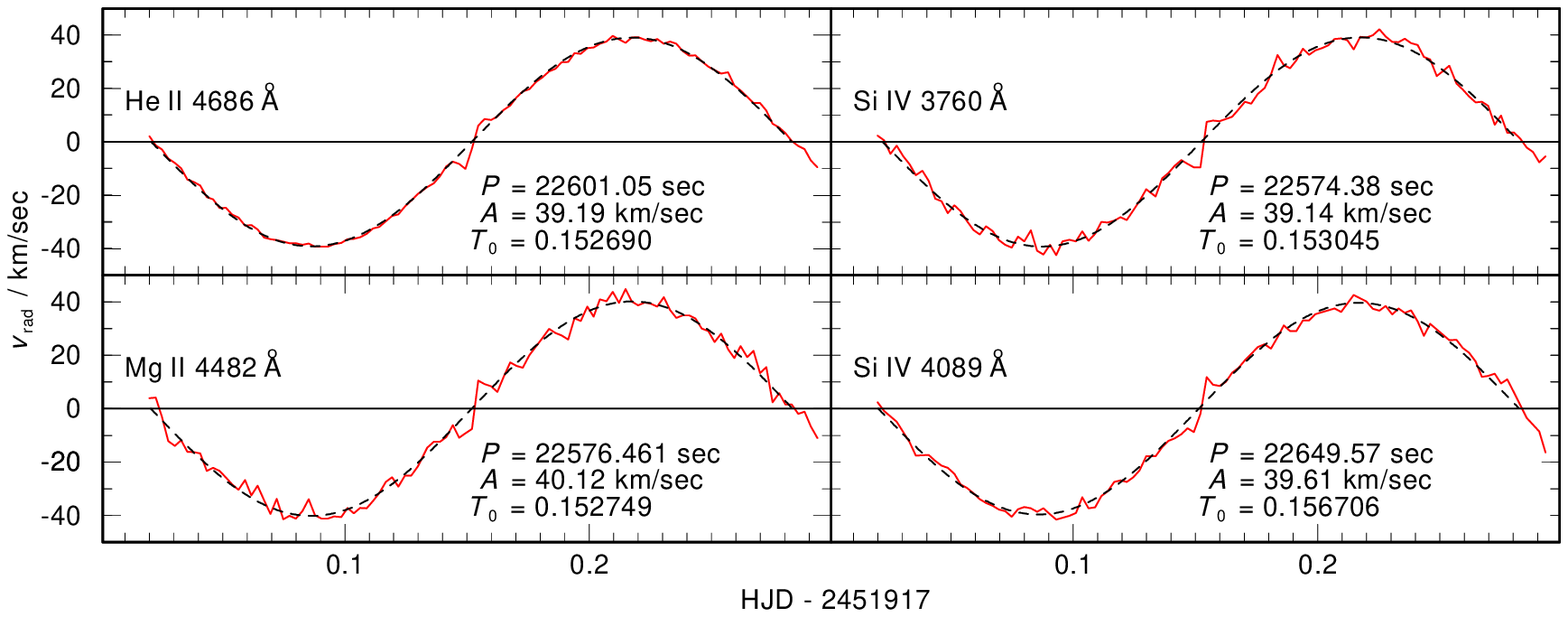}}
  \caption[]{Radial-velocity curve of \object{LB\,3459} measured from the individual lines.
             \Ionw{He}{2}{4686}, \Ionw{Mg}{2}{4482}, and \Ionww{Si}{4}{3760, 4089} have the sharpest observed
             line profiles (shown here) while the Balmer lines are much broader and the S/N ratio is
             significantly decreasing towards the higher series members \sA{coadd}.
             Note the velocity jumps close to $T_0$ which are the result of the transit
             of the cool companion \sK{trans}!
            }
  \label{orb}
\end{figure*}

In a recent spectral analysis (Rauch 2000) of the primary, based on high-resolution
CASPEC and IUE spectra, $M_1 =0.330\,\mathrm{M_\odot}$ has been derived from
comparison of \Teff\ and \logg\ to evolutionary tracks of post-RGB stars (Driebe \ea\,1998).
$M_2 = 0.066\,\mathrm{M_\odot}$ has subsequently been calculated from the system's mass function.
Thus, the secondary is formally a brown dwarf ($0.013 - 0.08\,\mathrm{M_\odot}$).

{\sc HHH} found that \object{LB\,3459} lies in that region of the Hertzsprung-Russell
diagram appropriate for binaries undergoing late case B 
(Roche lobe filled just before central helium ignition)
mass transfer (Iben \& Livio 1993) followed by a common-envelope (CE) phase. 

An alternative CE scenario has been studied by Livio \& Soker (1984). They found that in 
star--planet systems, when the star evolves into a red giant, the planet may survive
the CE phase when its intial mass is above $M_\mathrm{crit} \approx 0.02\,\mathrm{M_\odot}$.
Then the planet can accrete from either the stellar wind or directly from the giant's envelope.
Thus, it appears possible that the secondary of \object{LB\,3459} had formerly been a planet
which has transformed into a low-mass stellar component.

However, the mass-radius relation of \object{LB\,3459} derived from the mass function and 
the light-curve analysis (Kilkenny \ea\,1981) does not intersect with the empirical result of
Rauch (2000). The reason for this is unclear. Two possible reasons are 
that the theoretical models by Driebe \ea\,(1998) are not applicable to the case of
\object{LB\,3459} and the uncertainty for the photospheric parameters in Rauch (2000).

Since the analysis of Rauch (2000) was hampered by the relatively long exposure
times (1\,h and 2 -- 3\,h, respectively) and hence, a relatively large  
orbital velocity coverage (the observed line profiles were broadened by the star's 
rotation as well as by a smearing due to orbital motion within the observations),
it has been speculated that \logg\ from Rauch (2000) is somewhat too low. 
In order to make progress and to minimize the effects of orbital motion, 
spectra with short exposure times (180 sec) have been taken with UVES attached
to the ESO VLT (Kueyen) which cover a complete orbital period \sK{obs}.
From these spectra, the radial-velocity curve of \object{LB\,3459} is 
determined \sK{rad}. In \se{trans} we will have a close look onto the spectral
changes during the transit of the cool component. From those spectra with a minimum 
effect of orbital smearing ($v_\mathrm{rad}$ at maximum, \ab{dvrad}), we can 
then derive $v_\mathrm{rot}$ \sK{rot}.

\section{Observations}
\label{obs}

Two UVES spectra have been taken on Jan 6, 2001, with an exposure time of 180\,sec each. Then the run was
stopped although the night had been photometric. It was restarted on Jan 8, 2001, and
105 UVES spectra were taken in a partly cloudy night. The improvement by the short exposure time is shown in
\ab{dvrad}. 
We used a slit width of $1\farcs0$ projected at the sky and arrive at a resolving power $R \approx 48\,000$.
The spectra of Jan\,8 cover a complete orbital period of \object{LB\,3459}. 

The spectra were subject to the standard reduction provided by ESO.
The S/N ratio of the single spectra is about $20-30$ while from the results of the UVES exposure-time
calculator we had expected $\mathrm{S/N}=80-110$ at a resolution of 35\,000. In \ab{coadd} we show our co-added spectrum:
All 105 spectra (Jan 8) were shifted to the rest wavelength, then co-added and subsequently binned to $0.045\,\mathrm{\AA}$
(3 wavelength points). Although the S/N of the co-added UVES spectrum is about 200, no chemical 
element could be identified which had not been detected in other spectra before (Rauch 2000).

\section{Radial velocity}
\label{rad}

\begin{table}
\caption{Orbital period of \object{LB\,3459} determined from different datasets. 
The typical error in the period is 0.05 sec.}
\label{ptab}
\begin{tabular}{rr@{.}ll}
\hline
\noalign{\smallskip}
No\@. of measurements & \multicolumn{2}{c}{$P$ / sec} & data source\\
\hline       
\noalign{\smallskip}
105 & 22643&921 & Jan 8 \\  
107 & 22601&138 & Jan 6 + Jan 8 \\
131 & 22600&702 & Jan 6 + Jan 8 + {\sc hhh} \\
 24 & 22590&326 & {\sc hhh} \\
\hline
\end{tabular}
\end{table}

In the first step, the phase-dependent radial velocity is determined by fitting Lorentzians
to H\,$\beta$ -- H\,$10$, \Ionw{He}{2}{4686}, \Ionw{Mg}{2}{4482}, and \Ionww{Si}{4}{3760, 4089} in all 107 spectra.
Other lines have been checked but turned out to be useless: \Ionw{C}{2}{3921} (too weak),
\Ionww{Si}{4}{3921, 4116} (blends with H\,$\epsilon $ and \Ion{C}{4}, respectively),
and \Ionww{O}{3}{3755, 3758, 3740} (too weak).
During this procedure the smearing due to orbital motion during the exposure was neglected because the exposure
time \sK{int} is very short compared to the orbital period. 
Then, the derived velocity curves were fitted by sine curves in order to determine their periods, amplitudes, and $T_0$.

The orbital period has been determined from the complete data set because the 107 spectra provide a time line of
9 periods which increases the accuracy.
For the amplitude and $T_0$ determinations, we use only the 105 spectra of Jan 8 because the scatter in the two
spectra of Jan 6 increase the scatter in the results artificially. The times (including heliocentric correction) are the 
middles of the exposures. The results are summarized in \ab{orb}.

Since the observed \Ionw{He}{2}{4686} line is the sharpest line in the optical spectrum
(Rauch 2000), we achieve obviously \sA{orb} the best fit to a theoretical sine curve with the velocity curve 
measured from this line (the same line had been used by {\sc hhh}). 
\Ionw{Mg}{2}{4482} is also sharp but much weaker and the Balmer 
lines are much broader and the S/N is lower. The measured amplitude of the velocity
curve appears to be larger for the higher Balmer lines. This is clearly an effect of the
lower S/N. Thus, we decided to use only \Ionw{He}{2}{4686} for the period determination.
Moreover, {\sc hhh} provide 24 additional measurements of the radial-velocity curve of \object{LB\,3459} 
based on \Ionw{He}{2}{4686} which extends the time base up to 9692 periods. 
In \ta{ptab}, we summarize the results based on different datasets.

\begin{figure}[ht]
  \resizebox{\hsize}{!}{\includegraphics{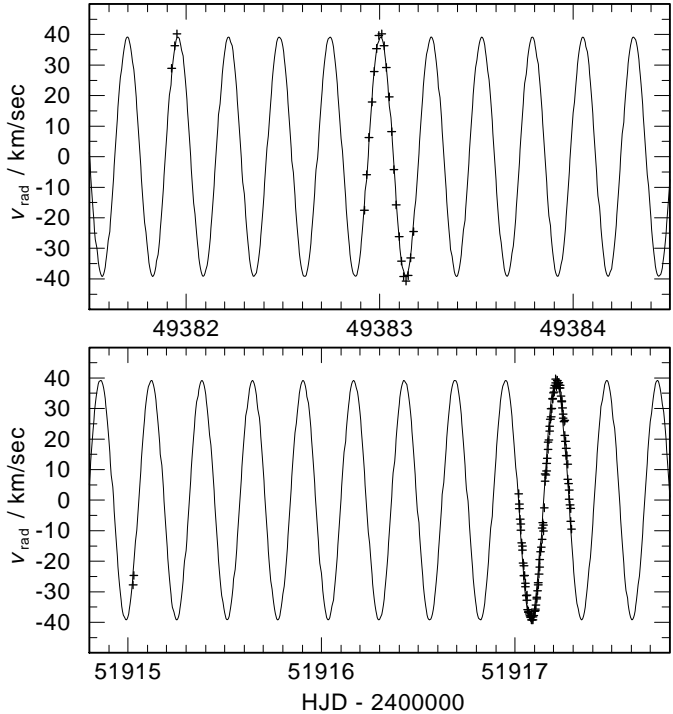}}
  \caption[]{Comparison of radial-velocity measurements (Top: {\sc hhh} data, bottom: our data)
             of \object{LB\,3459} with a sine curve
             calculated from our solution ($P=22\,600.702\,\mathrm{sec}$, $A=39.19\,\mathrm{km/sec}$,
             $T_0 = 51917.15269$). Note that the measurements cover the time from Jan 29, 1994, to
             Jan 8, 2001.
            }
  \label{sinfit}
\end{figure}

\begin{figure}[ht]
  \resizebox{\hsize}{!}{\includegraphics{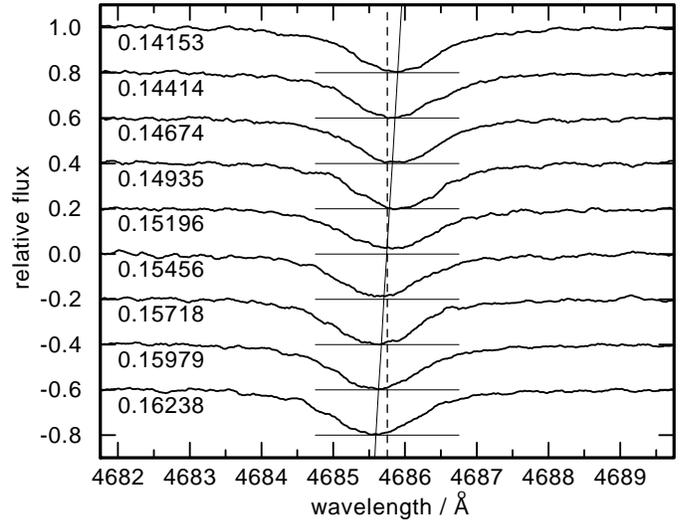}}
  \caption[]{The observed line profile of \Ionw{He}{2}{4686} in spectra taken during the transit
             of the secondary (times are given in HJD - 2451917). 
             The spectra are binned to a resolution of $0.1\,\mathrm{\AA}$. The central depression
             is obviously smaller at $T=0.15196$ ($T_0=0.15269$). The dashed line indicates
             the rest wavelength, the vertical curve is the wavelength shift calculated from
             the radial-velocity curve \sK{rad}, 
             the thin horizontal lines indicate the central depression outside the transit phase.
            }
  \label{radius}
\end{figure}
 
\begin{figure}[ht]
  \resizebox{\hsize}{!}{\includegraphics{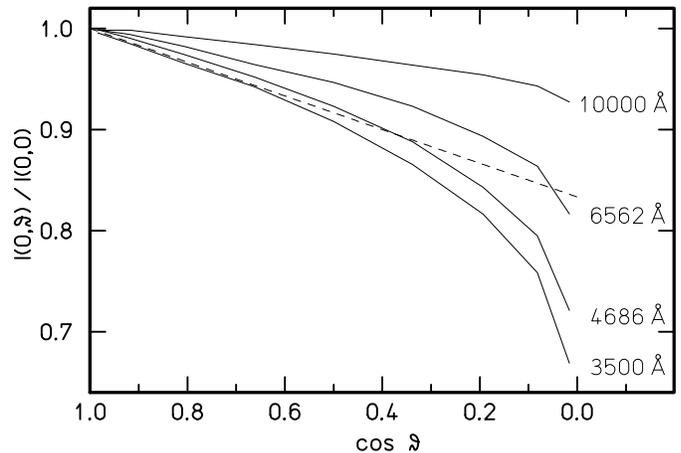}}
  \caption[]{Limb darkening in \object{LB\,3459} at four wavelengths as derived from our model atmosphere. 
             Note that a standard darkening law 
             (grey approximation)  
             $\frac{I_\mathrm{\nu}(0,\vartheta)}{I_\mathrm{\nu}(0,0)} = \frac{1+\beta_0 \cos\vartheta}{1+\beta_0}$
             (e.g\@. Uns\"old 1968, here calculated with $\beta_0 = 0.2$, dashed line) is only a poor approximation
             for values which are calculated from a more realistic line-blanketed NLTE model atmosphere.
            }
  \label{limb}
\end{figure}

From the result of the dataset of \Ionw{He}{2}{4686} which includes additionally the {\sc hhh} data \sT{ptab},
we adopt \mbox{$P=22\,600.702\spm 0.005\,\mathrm{sec}$} and \mbox{$A_1=39.19\spm 0.05\,\mathrm{km/sec}$} \sA{sinfit}.
The period is only 0.02\% longer than given by Kilkenny \ea\ (1991, 22\,597.0\,sec from photometric data of
27 eclipses). 
A higher radial-velocity amplitude of $A_1=40.8\spm 0.7\,\mathrm{km/sec}$ has been measured by {\sc hhh}
from their observations of \Ionw{He}{2}{4686}.

\section{The transit of the cool companion}
\label{trans}

The transit of the cool companion is clearly visible in the radial-velocity curves \sA{orb}.
An inspection of the observed \Ionw{He}{2}{4686} line profiles during this transit shows 
a decrease of the central depression close to $T_0$ \sA{radius}.

In order to simulate the phase-dependent variation of the line profile of \Ionw{He}{2}{4686},
we have set up a surface grid (with 25 azimuth and 25 altitude points) on the primary, and added up the spectra from 
these grid points with correct projected $v_\mathrm{rad}$ and consideration of limb darkening,
calculated from the final model atmosphere for \object{LB\,3459} \sA{limb}.

During the transit of the cool companion, the hidden parts of the primary's surface
do not contribute to the resulting spectrum \sA{transit}. 
It can be seen that the line profile during the transit is first redshifted, then blueshifted, as
can be expected. This can be seen as well in the radial-velocity curves \sA{orb}. 
At the begin and the end of the eclipse, the line profile appears deeper 
(Rossiter effect, Rossiter 1924).
In addition, at the central eclipse, the line profile is shallower compared to phases
shortly before and after. This is an effect of limb darkening, and it is qualitatively
reproduced by our model atmosphere. 
From the comparison of the \Ionw{He}{2}{4686} line profile at selected orbital phases 
($\varphi = 355.7\degr, 359.3\degr, 6.5\degr$), $r_1/r_2$ appears to be at its upper limit 
\sA{orb}. It is worthwhile to note, that
a phase-resolved spectroscopic study of line-profile variation gives the radius ratio $r_1/r_2$ 
independently from light-curve analysis --- if the quality of the spectra and the phase resolution are sufficient.

We have produced a short animation (available at the CDS)
which shows all our observed spectra compared to the simulation
as well as a view onto the orbital movement of the binary. In \ab{movie} we show 
a snapshot of this animation.

\begin{figure}[ht]
  \resizebox{\hsize}{!}{\includegraphics{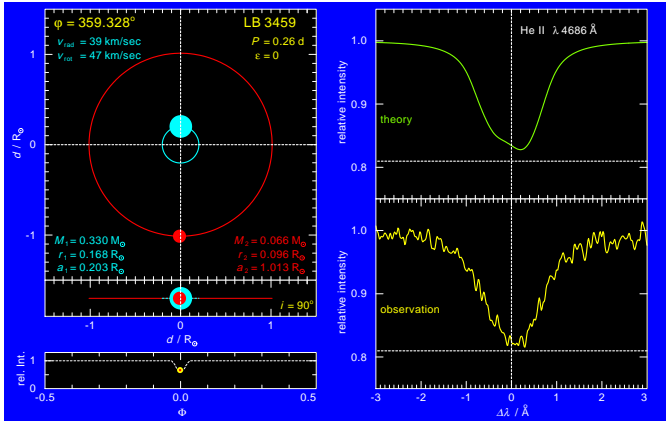}}
  \caption[]{Phase-dependent spectral variation of the \Ionw{He}{2}{4686} line profile. A snapshot taken
             from our animation. Left panels: View onto the binary system LB\,3459 from the top and from the
             side (as seen from Earth, $i = 90\degr$), 
             at the bottom, the normalized lightcurve of the system is shown.
             Right: Our UVES spectra (bottom, smoothed by a
             Gaussian of 0.05\,\AA\ FWHM) compared to the synthetic spectrum (see text for details).}
  \label{movie}
\end{figure}

\begin{figure*}[ht]
  \resizebox{\hsize}{!}{\includegraphics{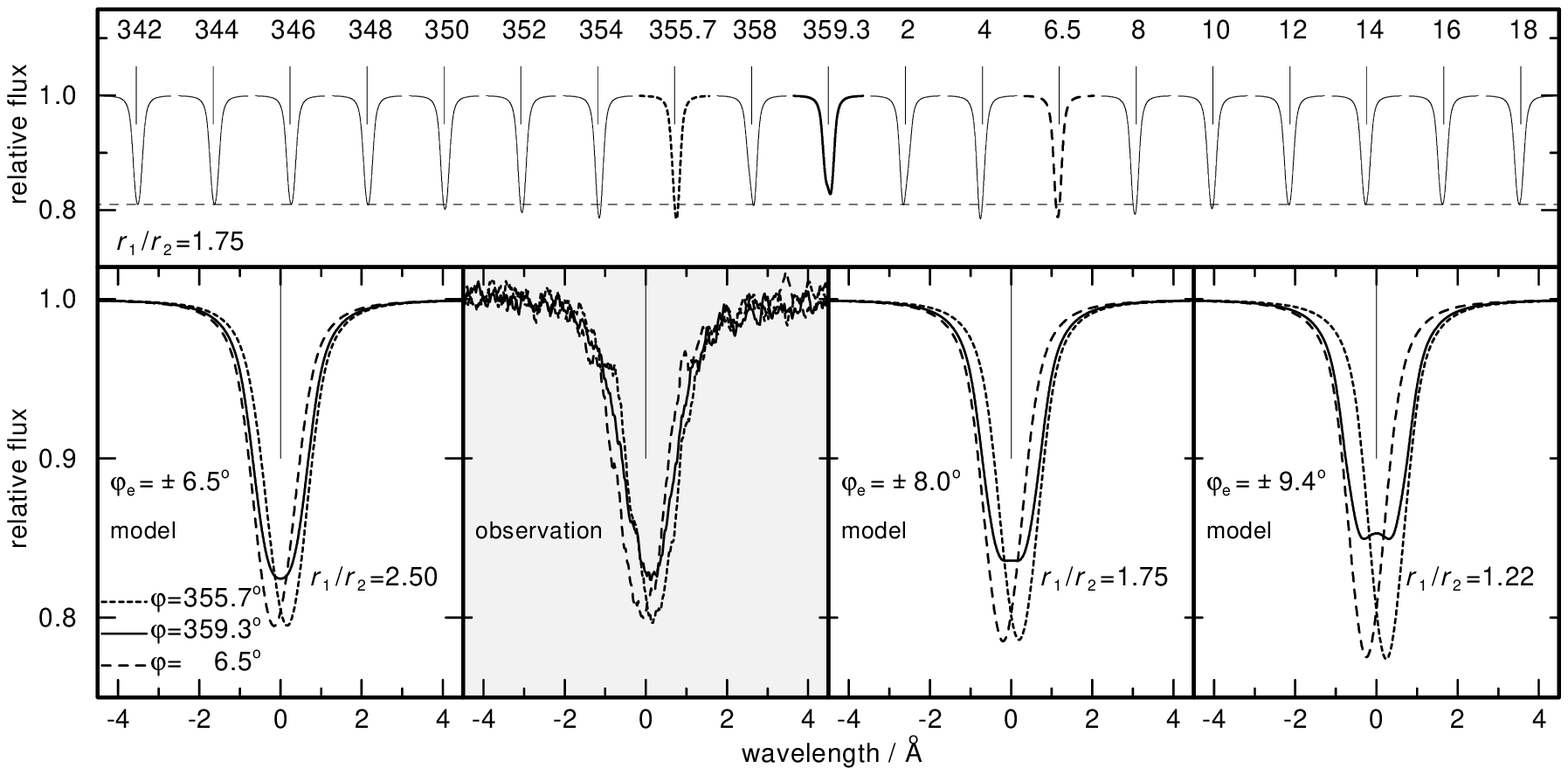}}
  \caption[]{Top: Predicted phase-dependent variation ($\varphi = 342\degr - 18\degr$)
             of the theoretical \Ionw{He}{2}{4686} line profile during the transit of the secondary. 
             The dashed horizontal line indicates the central depression outside the transit phase.
             Bottom: Comparison of theoretical \Ionw{He}{2}{4686} line profiles at
             three different radius ratios, 
             $r_1/r_2 = 2.50, 1.75, 1.13$ with $r_1 = 0.168 \spm 0.03\,\mathrm{R_\odot}$
             which are corresponding to the error limits of $r_1$ and $r_2$ (Rauch 2000),
             with the observation (second panel, data binned to $0.1\,\mathrm{\AA}$). 
             $\varphi_e$ gives the angular limits where the primary is eclipsed.
             }
  \label{transit}
\end{figure*}

\section{Rotational velocity of the sdOB primary}
\label{rot} 

The maximum change of $v_\mathrm{rad}$ during a 180\,sec-exposure of \object{LB\,3459}
is $\Delta v_\mathrm{rad} = 1.958\,\mathrm{km/sec}$ \sA{dvrad} and thus, the ``line-broadening'' 
effects of smearing due to orbital motion are small. Since the S/N of a single of our UVES spectra
is about 20 (in average), we decided to co-add all spectra \sA{coadd} in order to achieve the maximum
S/N ratio. The alternative, to select only those spectra where $\Delta v_\mathrm{rad}$ is below a certain
limit, e.g\@. smaller than $0.5\,\mathrm{km/sec}$, and to reduce the orbital smearing would reduce 
the percentage of spectra to co-add to 15\,\% --- resulting in a much lower S/N.

Since the resolution of the spectra is better than 0.1\,\AA\ (at the position of \Ionw{He}{2}{4686}),
we checked whether it is necessary to consider the fine-structure splitting of \Ionw{He}{2}{4686}.
In \ab{fine} we show that this results in a wider line profile and cannot be
neglected, otherwise the helium abundance is strongly overestimated. 

\begin{figure}[ht]
  \resizebox{\hsize}{!}{\includegraphics{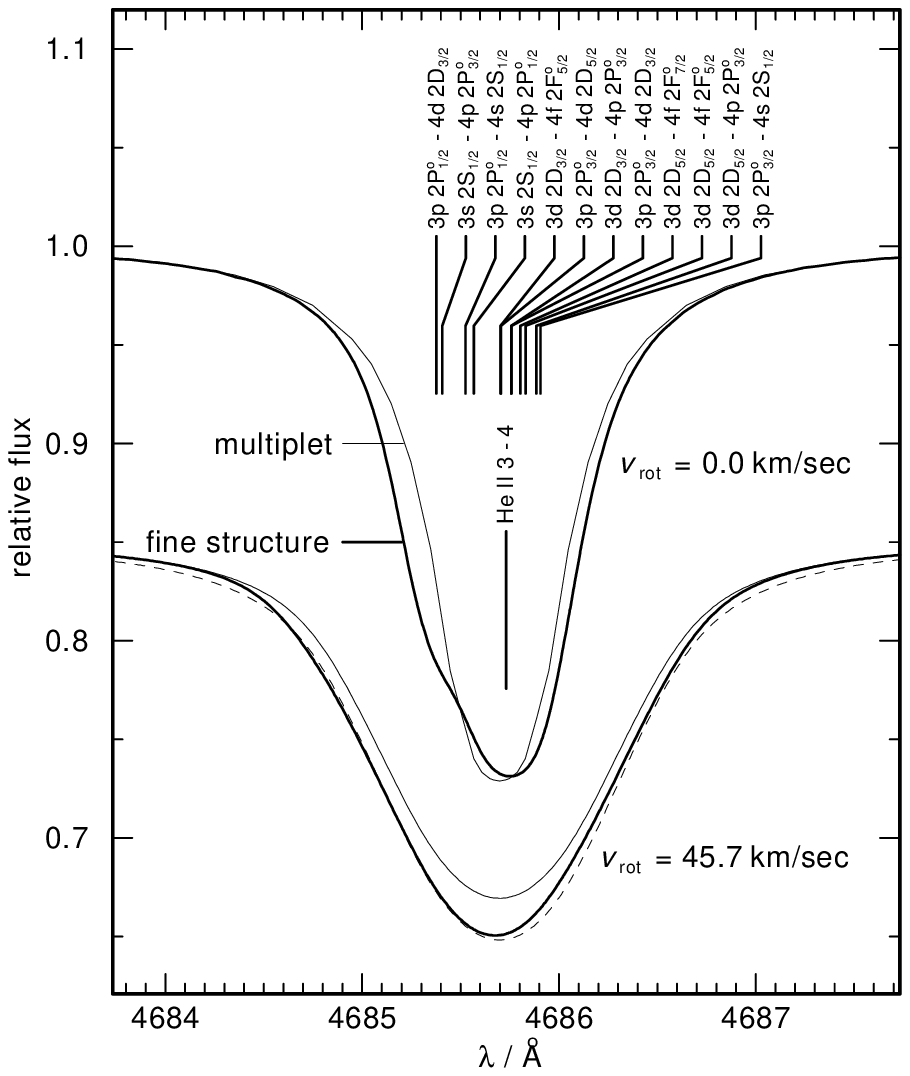}}
  \caption[]{Theoretical line profiles of \Ionw{He}{2}{4686} calculated from models with
             \Teffw{42}, \loggw{5.2}, and \ratiow{He}{H}{0.0008} (Rauch 2000). The thick lines show
             the profile which is calculated with fine structure splitting (components are
             indicated). For the thin lines, \Ionw{He}{2}{4686}
             is calculated as one multiplet. Note that the deviation is larger than the spectral
             resolution (0.1\,\AA) of our observation. At the bottom, the profiles are convolved
             with a rotational profile ($v_\mathrm{rot} = 45.7\,\mathrm{km/sec}$) and compared to
             a theoretical line profile (dashed, without fine structure splitting) calculated from model
             with a 50\% higher helium abundance (\ratiow{He}{H}{0.0012})
            }
  \label{fine}
\end{figure}

We calculated a grid of H+He composed NLTE model atmospheres in the relevant parameter range, i.e\@.
\Teffw{40 - 45} (in steps of 1\,kK), \loggw{4.95 - 5.55} (0.05, cgs), 
\ratiow{He}{H}{0.0001 - 0.0020} ($0.0001$, number ratios), and calculated theoretical
line profiles of \Ionw{He}{2}{4686}. We performed a $\chi^2$ test (wavelength, flux level, 
\Teff, \logg, \ratio{He}{H}, and $v_\mathrm{rot}$) in order to determine the rotational velocity.
The lowest reduced $\chi^2$ value (0.44) is found at \Teffw{44}, \loggw{5.4}, \ratiow{He}{H}{0.0007}, 
and $v_\mathrm{rot} = 43\,\mathrm{km/sec}$. However, Rauch (2000) has determined \Teffw{42\spm 1} from
ionization equilibria (\Ion{He}{1}\,/\,\Ion{He}{2}, \Ion{C}{3}\,/\,\Ion{C}{4}, \Ion{N}{3}\,/\,\Ion{N}{4}/\,\,\Ion{N}{5},
\Ion{O}{4}\,/\,\Ion{O}{5}) which are very sensitive indicators for \Teff, and \loggw{5.21\spm 0.1} from the
broad hydrogen Balmer lines. Since the error ranges are small, we have to keep these values fixed.
Then, $v_\mathrm{rot} = 47\,\mathrm{km/sec}$ and \ratiow{He}{H}{0.0008} yield the best fit. 
In \ab{vrot} the co-added spectrum is compared with theoretical line profiles of \Ionw{He}{2}{4686}
which are convolved with rotational profiles with different rotational velocities and
subsequently with a Gaussian of 0.1\,\AA\ (FWHM) in order to account for the instrumental profile.
From this comparison we have estimated an error range of $\spm 5\,\mathrm{km/sec}$ for $v_\mathrm{rot}$
in this analysis.

The \ratio{He}{H} ratio is the same 
as found by Rauch (2000) who used CES spectra with a
resolution of 0.1\,\AA. Since the exposure time of the CES spectra had been
3600\,sec, the orbital smearing \sA{dvrad} and the lower S/N of about 20 made the fine-structure splitting
\sA{fine} unimportant.

\begin{figure}[ht]
  \resizebox{\hsize}{!}{\includegraphics{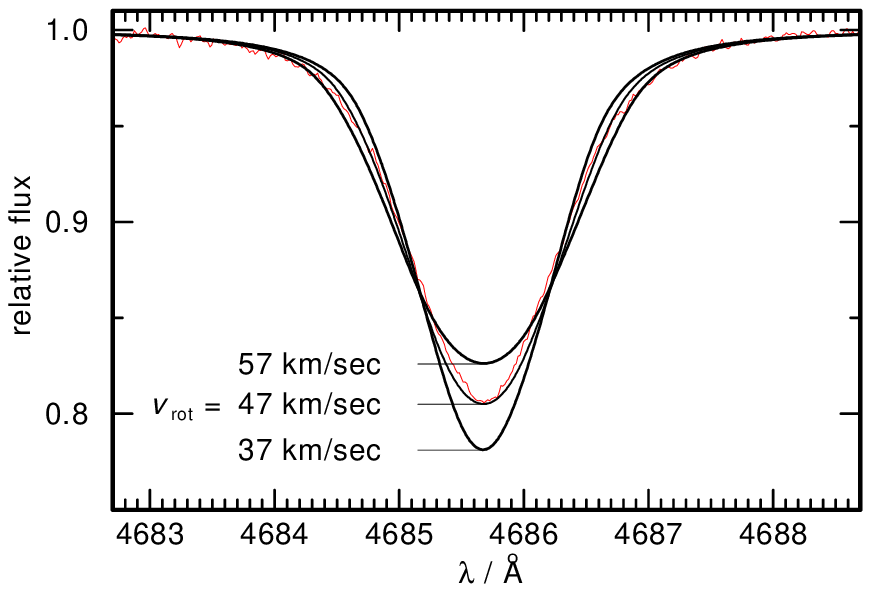}}
  \caption[]{Comparison of theoretical line profiles (\Teffw{42}, \loggw{5.2}, \ratiow{He}{H}{0.0008}) 
             of \Ionw{He}{2}{4686} to the co-added
             UVES spectrum. At $v_\mathrm{rot} = 47\,\mathrm{km/sec}$
             we find a good agreement. The error range can be estimated to $\spm 5\,\mathrm{km/sec}$.
             }
  \label{vrot}
\end{figure}

\section{Dimensions of \object{LB\,3459}}
\label{dim}

With the orbital period $P$ and the radial velocity amplitude $A_1$ we can calculate
the circumference $U_1$ and radius $a_1$ of the primary's orbit. From the relation
$a_1M_1=a_2M_2$ we can derive parameters for the secondary. The components' separation
is $a=a_1+a_2$. Kilkenny \ea\ (1979) have given formulae for the radii of the 
components: $r_1=0.138a$ and $r_2=0.079a$ derived from the light curve. 
The calculated values are summarized in \ta{dimt}.
In \ab{dimpl} the dimensions of \object{LB\,3459} are illustrated.

\begin{figure}[ht] 
  \resizebox{\hsize}{!}{\includegraphics{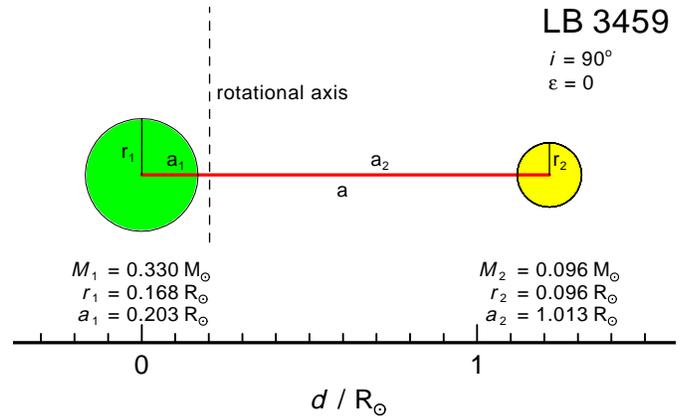}}
  \caption[]{Dimensions of \object{LB\,3459}.}
  \label{dimpl}
\end{figure}

\begin{table}
\caption{Orbital parameters of \object{LB\,3459}. For $r_1$ and $r_2$, see text for details.}
\label{dimt}
\begin{tabular}{rr@{.}lr@{.}lc}
\hline
\noalign{\smallskip}
$P$     &   22600&702    & \spm\ 0&005          & sec                   \\
$T_0$   & 2451917&152690 & \spm\ 0&000005       & HJD                   \\
\hline
\noalign{\smallskip}
$A_1$   &      39&19     & \spm\ 0&05           & km/sec                \\
$a_1$   &       1&4097   & \spm\ 0&0021         & $10^{10}\,\mathrm{cm}$ \\
$U_1$   &       8&8572   & \spm\ 0&0130         & $10^{10}\,\mathrm{cm}$ \\
        &       0&2025   & \spm\ 0&0019         & $\mathrm{R}_\odot$    \\
\hline
\noalign{\smallskip}
$A_2$   &     195&95     & \spm\ 6&55           & km/sec                \\
$a_2$   &       7&0485   & \spm\ 0&2321         & $10^{10}\,\mathrm{cm}$ \\
        &       1&0127   & \spm\ 0&0335         & $\mathrm{R}_\odot$    \\
$U_2$   &       4&4287   & \spm\ 0&1458         & $10^{11}\,\mathrm{cm}$ \\
\hline
\noalign{\smallskip}
$a$     &       8&4582   & \spm\ 0&2342         & $10^{10}\,\mathrm{cm}$ \\
        &       1&2153   & \spm\ 0&0354         & $\mathrm{R}_\odot$    \\
\hline
\noalign{\smallskip}
$r_1$   &       1&1672   & \spm\ 0&0323         & $10^{10}\,\mathrm{cm}$ \\
        &       0&1678   & \spm\ 0&0047         & $\mathrm{R}_\odot$    \\
$r_2$   &       0&6682   & \spm\ 0&0184         & $10^{10}\,\mathrm{cm}$ \\
        &       0&0960   & \spm\ 0&0003         & $\mathrm{R}_\odot$    \\
\hline
\end{tabular}
\end{table}

We calculate $r_1=0.1678 \spm 0.03\,\mathrm{R}_\odot$ for the primary.
This is about $1/\sqrt{2}$ smaller than $r_1=0.236\,\mathrm{R}_\odot$ derived by Rauch (2000)
from $r_1 = \sqrt{G M_1/g}$. Although $M_1$ is determined by comparison of \Teff\ and
\logg\ to theoretical models \sK{int}, $g$ itself is the main error source in this formula ---
a change of 0.3 in \logg\ (factor 2 in $g$) would change $M_1$ only by a few percent.
Since all values of \ta{dimt} have been calculated straightforward, the errors are mainly
propagation errors. This may be a hint for a higher $g$ than determined by Rauch (2000).

However, the velocity value of $v_\mathrm{rot} = 45.7\,\mathrm{km/sec}$ for bound rotation
following Rauch (2000), i.e\@. calculated with $r_1=0.236\,\mathrm{R}_\odot$ and $P=22\,600.702\,\mathrm{sec}$
would be smaller by $1/\sqrt{2}$ as well. Then, the present rotational velocity is much higher
than this value.
For this paper, we decided to adopt $r_1=0.168\,\mathrm{R}_\odot$  and $r_2=0.096\,\mathrm{R}_\odot$. 
However, the origin of this discrepancy needs further investigation.

\section{Results and discussion}

We have determined the orbital period of \object{LB\,3459} from the newly measured radial velocity
curve in good agreement with previous values of Kilkenny \ea\ (1991). Together with data from {\sc hhh}
we could strongly reduce the error limits \sT{dimt}.

We derived a rotational velocity of the primary of \object{LB\,3459} of
$v_\mathrm{rot} = 47\spm 5\,\mathrm{km/sec}$ which is in its error limits in agreement with 
$v_\mathrm{rot} = 34\spm 10\,\mathrm{km/sec}$ found by Rauch (2000).
For $r_1=0.236\,\mathrm{R}_\odot$, this is equivalent to a bound rotation. If the radius is $r_1=0.168\,\mathrm{R}_\odot$
\sK{dim}, then the primary rotates about 40\% faster than bound.

The radius of the cool component is $r_2 = 6.682\cdot10^9\,\mathrm{cm}$ which is almost the same size
like Jupiter ($r = 7.143\cdot10^9\,\mathrm{cm}$) but its mass is about 1/15\,M$_\odot$ which is much higher
than Jupiter's mass (1/1047\,M$_\odot$). Thus, from its present mass, the secondary of \object{LB\,3459}
lies formally within the brown-dwarf mass range ($0.013 - 0.08\,\mathrm{M_\odot}$). 
The numerical simulations of star--planet systems presented by Livio \& Soker (1984)
show that a critical mass exists above which it is possible that a planet with $M_2 > M_\mathrm{crit}$
gains mass during a CE phase. Due to a simplified treatment of accretion and evaporation (Iben \& Livio 1993) in
the work of Livio \& Soker (1984), improved calculations of star--planet systems are highly desirable
to verify the idea of Rauch (2000), that the secondary of \object{LB\,3459} is a former planet.

From the comparison of \Ionw{He}{2}{4686} to the co-added UVES spectrum \sA{vrot},
the helium abundance of \ratiow{He}{H}{0.0008\spm 0.0002} found by Rauch (2000)
has been confirmed by the use of spectra with higher S/N and higher resolution.
Moreover, the rotational velocity found here is very close to the
value of $v_\mathrm{rot} = 45\,\mathrm{km/sec}$ which had been adopted by Rauch (2000) in
his analysis. 

Although the quality of the co-added UVES spectrum is much higher than previously available
spectra, no elements have newly been detected in it. 

The reason for the discrepancy between primary and secondary masses derived from spectral
analysis and those from the radial-velocity and the eclipse curves \sK{int} is still 
unclear --- although some problems in the spectral analysis due to uncertain rotational broadening 
and smearing due to orbital motion could be ruled out. There is a hint, that the surface
gravity $g$ may be slightly higher than determined by Rauch (2000). If true, this would 
solve the problem with the discrepancy. Unfortunately, significant effects
of changes in $g$ are detectable only in the outer line wings where the data reduction of the 
broad Balmer lines in the echelle spectra is not very accurate.

To summarize, there is no reason to perform a new spectral analysis of \object{LB\,3459} based
on the new optical UVES spectra. However, it is highly desirable to obtain high-resolution and 
high-S/N spectra in the UV/FUV in order to improve the determination of \Teff\ based on 
ionization equilibrium of metals, to improve the determination of \logg\ by the evaluation of the higher
members of the Lyman series, and to identify other elements beside C, N, O, Si, Mg, Fe,
and Ni in order to prove likely diffusion effects found by Rauch (2000).

\begin{acknowledgements}
The UVES spectra used in this analysis were obtained as part of an ESO Service Mode run,
proposal 66.D-1800.
We like to thank Stefan Dreizler and J\"orn Wilms for their assistance in IDL programming,
and Christian Karl for help in the data reduction.
Computations were carried out on CRAY computers of the Rechenzentrum der Universit\"at Kiel.
This research has made use of the SIMBAD Astronomical Database, operated at CDS, Strasbourg, France.
This research was supported by the DLR under grants 50\,OR\,9705\,5 and 50\,OR\,0201.
\end{acknowledgements}

\end{document}